\documentclass[acmsmall]{acmart}

\AtBeginDocument{%
  \providecommand\BibTeX{{%
    \normalfont B\kern-0.5em{\scshape i\kern-0.25em b}\kern-0.8em\TeX}}}

\setcopyright{rightsretained}
\copyrightyear{2021}
\acmYear{2021}
\acmDOI{}
\acmISBN{} 
\acmConference[2nd KDD Workshop on Data-driven Humanitarian Mapping]{SIGKDD '21: 2nd KDD Workshop on Data-driven Humanitarian Mapping: Harnessing Human-Machine Intelligence for High-Stake Public Policy and Resiliency Planning}{August 15}{2021}

\begin{document}

\title{Estimating Active Cases of COVID-19}

\author{Javier Álvarez}
\author{Carlos Baquero}
\author{Elisa Cabana}
\author{Jaya Prakash Champati}
\author{Antonio Fernández Anta}
\author{Davide Frey}
\author{Augusto García-Agúndez}
\author{Chryssis Georgiou}
\author{Mathieu Goessens}
\author{Harold Hernández}
\author{Rosa Lillo}
\author{Raquel Menezes}
\author{Raúl Moreno}
\author{Nicolas Nicolaou}
\author{Oluwasegun Ojo}
\author{Antonio Ortega}
\author{Jesús Rufino}
\author{Efstathios Stavrakis}
\affiliation{%
  \institution{CoronaSurveys Team, \url{https://coronasurveys.org/team/}}
  \country{Spain}
  }
\author{Govind Jeevan}
\author{Christin Glorioso}
\affiliation{%
  \institution{DICE Institute, Pathcheck Foundation}
    \country{USA}
  }
  
\date{}

\begin{abstract}
    Having accurate and timely data on confirmed active COVID-19 cases is challenging, since it depends on testing capacity and the availability of an appropriate infrastructure to perform tests and aggregate their results. In this paper, we propose methods to estimate the number of active cases of COVID-19 from the official data (of confirmed cases and fatalities) and from survey data. We show that the latter is a viable option in countries with reduced testing capacity or suboptimal infrastructures. 
\end{abstract}

\settopmatter{printacmref=false}
\authorsaddresses{}

\maketitle
\renewcommand{\shortauthors}{Álvarez et al.}

\section{Introduction}

The COVID-19 disease epidemic caused by the SARS-CoV-2 virus started having a deep impact in societies across the world as the virus spread globally in early 2020. Not surprisingly, during its initial months, the knowledge about virus properties and the ability to test for its presence or correctly identify symptoms was limited \cite{maxmen2020much}. 
These factors, when coupled with the fast spread of the pandemic made it very challenging to obtain accurate estimates of the numbers of individual cases. 
This problem affected the estimation of a number of important metrics, including the number of daily cases, the cumulative number of cases, or the proportions of active and contagious cases at a given time and place.

In this article, we address these limitations and propose several approaches that try to predict the true magnitude of actual cases. Specifically, we aim to assess the number of active cases at a given point in time, i.e., how many people are sick with COVID-19. The first class of approaches depend on the accuracy of the official reporting of COVID-19 cases and deaths \cite{jhu-crc,worldometers}, which can be an issue in some regions. The second class of approaches is survey-based and built on data collected by the CoronaSurveys project \cite{coronasurveys-web}, and by the University of Maryland and Carnegie Mellon University with the support of Facebook \cite{FB-survey}. We observe that in some countries (like Greece or Portugal) the most recent estimates from official data and from surveys match. However, in other countries (like Brazil or India) the surveys data gives a much larger number of active cases, possibly due to under reporting in the official data. The conclusion is that estimates via surveys is a viable option to estimate active number of cases in countries with reduced testing capacity or suboptimal reporting infrastructures. 


\section{Estimating the number of active cases from official data}
\label{s-official-data}

Soon after the genetic sequence of the virus was made public, in early 2020, it became possible to detect the virus  from biological samples by means of RT-PCR (reverse transcription-polymerase chain reaction), a technique that detects fragments of viral particles. Health authorities started collecting daily numbers of detected cases and reporting them, as well as the numbers of fatalities attributed to COVID-19. Several data aggregators of worldwide case numbers appeared, notably those at Johns Hopkins University \cite{jhu-crc} and Oxford \cite{OxCGRT}.

\subsection{Estimating active cases from daily data}

The simplest approach to estimate active cases consists in using the official numbers of confirmed cases, fatalities, and recovered cases that have been collected in several data repositories \cite{jhu-crc, ECDC, OxCGRT}. The official number of active cases can hence be computed by subtracting the number of fatalities and recovered from the number of confirmed cases. In this paper, we use the number of active cases reported by Johns Hopkins University \cite{jhu-crc} based on the above computation (labelled as \emph{JHU active} in the plots below). 


Unfortunately, the number of recovered cases is not reliable or available in all countries (e.g., Greece and Madrid), so in order to have another estimation of the official number of active cases we use a second approach: we assume that each reported case is active for a fixed period, which is the median number of days COVID-19 cases are active. We label this estimate as \emph{Confirmed} in the plots below.
To define the usual duration of a COVID-19 case, we consider the CDC Morbidity and Mortality Weekly Report~\cite{MMWR2020}. It states that ``(65\%) reported that they had returned to their usual state of health a median of 7 days (IQR = 5–12 days) from the date of testing'' and that ``the median number of days respondents reported feeling unwell before being tested for SARS-CoV-2 was 3 (IQR = 2–7 days)''. From these values we establish a reference of 10 ($3+7$) days of active case duration (from onset to recovery) as covering the majority of cases. The same active period length is proposed by Singanayagam et al.~\cite{singanayagam2020duration}.

\subsection{Estimating active cases from cumulative data}
\label{s-fatalities}

In this section we show how the official cumulative number of cases and fatalities can be used to estimate daily new cases, and from this value the active cases. By default, we will assume that the presented techniques are applied while immunity is mostly driven by natural infection, as is still the case in June 2021 for most low- and medium-income countries (LMIC).

\vspace{-0.5em}

\subsubsection{Tracking the cumulative number of cases}

Information on the cumulative number of detected cases in a given region is of particular relevance when trying do determine the proportion of population that had contact with the disease. We will show that combining this value with the cumulative number of COVID-19 deaths can help estimating the true number of cases in each region of concern. 
On March 13 2020 a news article in Nature \cite{maxmen2020much} brought attention to the high level of under detection of COVID-19 cases, and suggested the use of estimates of the case fatality rate ($\textsf{CFR}$, at that time estimated around $1\%$) to infer the likely number of cases. 
Also in March 2020 The Centre for the Mathematical Modelling of Infectious Diseases (CMMID) at the London School of Hygiene \& Tropical Medicine (LSHTM) made available a more detailed estimation of the level of under-detection of COVID-19 symptomatic cases
\cite{russel2020using,Russell2020.07.07.20148460}.
This study took into account the delay distribution from symptoms to death and used a more precise estimate of the $\textsf{CFR}$, from an ongoing study with Wuhan data, that established it around $1.38\%$ \cite{Verity2020}. 

Following the technique in \cite{Russell2020.07.07.20148460}, it is possible to estimate the actual cumulative number of symptomatic cases at time $t$ by looking at the reported cumulative number of cases and cumulative number of deaths, and calculating the apparent $\textsf{CFR}_t$. If this $\textsf{CFR}_t$ is higher than the baseline $\textsf{CFR}_b = 1.38\%$, this is an indication of under-detection of cases. The actual number of cumulative cases can be obtained by multiplying the cumulative reported cases by $\frac{\textsf{CFR}_t}{\textsf{CFR}_b}$. In \cite{DBLP:journals/corr/abs-2005-12783}, presented at the KDD 2020 Workshop on Data-driven Humanitarian Mapping, we provide a detailed description of this estimation technique, and show that it matched independent serology-based prevalence surveys, when considering Spanish data \cite{pollan2020prevalence}.   

\vspace{-0.5em}

\subsubsection{Daily cases - EpiCurve}

From the estimate of the true number of cumulative cases, we estimate the daily number of new cases (i.e., the EpiCurve) before we derive the number of active cases in each day.
Let $D=(d_1,\ldots,d_n)$ be the series of daily new cases and $C=(c_1,\ldots,c_n)$ the respective cumulative number of cases, then 
$d_i = c_i - c_{i-1}$ with $c_0=0$. 
If the $D$ series does not have negative values, then the corresponding series $C$ is non-decreasing (it is a rare event, but some countries have resorted to negative values in daily cases and/or deaths to correct past misreporting). 

As we know, from the previous section, the $D$ and $C$ series can under-estimate the true number of cases. The $\textsf{CFR}$ based techniques that we developed in \cite{DBLP:journals/corr/abs-2005-12783} provide a series  $\hat{C}$ that estimates the true number of cumulative cases. However, this series is not always non-decreasing, and needs some processing in order to estimate the likely true number of daily cases. 
%
Our target is to derive the series of estimated daily cases $\hat{D}$. 
To this end, we consider $\tilde{C}=(\tilde{c}_1, \ldots, \tilde{c}_n)$ where $\tilde{c}_i = \mathsf{max}(\hat{c}_1,\ldots,\hat{c}_i)$, so that the cumulative series $\tilde{C}$ is always non-decreasing. Now, we can define $\hat{D}$ such that $\hat{d}_i = \tilde{c}_i - \tilde{c}_{i-1}$ with $\tilde{c}_0 = 0$.

As an additional reference we also define a very simple estimator of daily cases $\hat{\mathcal{D}}$ that only takes into account the daily number of reported deaths. Let $F=(f_1,\ldots,f_n)$ be the series of daily fatalities attributed to COVID-19. We define $\hat{\mathcal{D}}=(\hat{d}_1,\ldots,\hat{d}_n)$ such that $\hat{d}_t = \frac{f_{t+13}}{0.0138}$, where $t$ is a day of reference and $t+13$ is a day 13 days later. That is, this considers the number of deaths in a given day and uses it to infer how many cases likely occurred 13 days before, by using as reference the Wuhan $\textsf{CFR}_b=1.38\%$. The delay of 13 days is the median time from onset of symptoms to death, estimated from the CDC COVID-19 Pandemic Planning Scenarios \cite{CDC-planning}. The disadvantage of this technique is that it cannot predict the daily cases in the most recent 13 days, while the advantage is that it only requires an accurate reporting of COVID-19 deaths, and does not need the series of detected cases. 

\vspace{-0.5em}

\subsubsection{Active cases}

In the previous section we have defined two estimates of daily new cases:   $\hat{D}$ \--- based on cases and deaths; and $\hat{\mathcal{D}}$ \--- only based on deaths. 
%
%
%
From them we derive two series of active cases: \textsf{CCFR} \--- derived from $\hat{D}$ and \textsf{CCFR Fatalities} \--- derived from 
$\hat{\mathcal{D}}$ (we use \textsf{CCFR} and \textsf{CCFR Fatalities} as labels in the plots below). Both series are obtained by considering each daily case as active the next 10 days (as we presented above), and letting each of them  expire after that time, so that they no longer contribute to subsequent estimates of active cases. This transformation also helps smoothing the data, since cumulative cases combine information from multiple days and reduce the effect of daily variations.
Note that both estimates are of \emph{symptomatic} active cases, since we use the $\textsf{CFR}$ to obtain them.

\section{Estimating active cases via surveys}
\label{s-survey-data}

The show now methodologies for deriving estimates of active cases from survey studies \cite{covidcast, garcia2021estimating,fernandez2021coronasurveys}.

\subsection{Estimating directly from symptoms}

Since the spring of 2020 Facebook has been collaborating in the context of their program Data for Good \cite{FB-survey} with CMU and U. of Maryland promoting surveys created by these two institutions \cite{kreuter2020partnering}. Both surveys collect thousands of daily responses to their surveys, in which participants report their symptoms, their habits, vaccination status, etc. The Delphi Group at Carnegie Mellon University U.S. COVID-19 Trends and Impact Survey  in partnership with Facebook (Delphi US CTIS) \cite{covidcast} collects around 50,000 responses daily in the USA, while University of Maryland Social Data Science Center Global COVID-19 Trends and Impact Survey  in partnership with Facebook (UMD Global CTIS) \cite{UMDsurvey} collects more than 100,000 responses daily from many countries in the world. For this article we have used the individual responses of UMD Global CTIS to estimate the total number of active cases using both direct and indirect survey questions. 

We observed that the set of responses to the UMD Global CTIS has to be curated. As a first approach,
we remove all responses that declare to have all 12 symptoms in the last 24 hours (fever, cough, difficulty breathing, fatigue, stuffy or runny nose, aches or muscle pain, sore throat, chest pain, nausea, loss of smell or taste, eye pain, headache). Additionally,
we remove the responses with high values is certain questions\footnote{We use 100 as the threshold to consider that a value is high. From the data we have observed that 100 is a conservative value that essentially filters out unrealistic responses, like those reporting millions.}
\begin{itemize}
    \item Declare at least one symptom and give a high value to the question ``For how many days have you had at least one of these symptoms?"
    \item Answer \emph{Yes} to question ``Do you personally know anyone in your local community who is sick with a fever and either a cough or difficulty breathing?" and give a high value to the question ``How many people do you know with these symptoms?"
    \item Give a high value to the question ``How many people slept in the place where you stayed last night (including yourself)?"
\end{itemize}

%
%

After filtering outliers, each survey response is considered to belong to a COVID-like illness active case (\textsf{UMD CLI} in the plots below) if
it declares the following combination of symptoms: fever, along with cough, shortness of breath, or difficulty breathing
(this is a combination used in the survey and the API \cite{UMDsurvey}). Similarly,
the response is considered to belong to a COVID-like illness according to the WHO (\textsf{UMD CLI WHO} in the plots) if it declares the
following symptoms: fever, cough, and fatigue 
\cite{who-faq}. Obviously, \textsf{UMD CLI} and \textsf{UMD CLI WHO} are estimates of the number of symptomatic active cases.

\subsection{Estimating with indirect reporting}


In \cite{garcia2021estimating,fernandez2021coronasurveys} we present the CoronaSurveys project, and show how it is possible to estimate the cumulative incidence of COVID-19 with indirect reporting and the Network Scale-up Method (NSUM) \cite{bernard1991estimating,laga2021thirty}. In the CoronaSurveys poll we have the following questions relative to a preselected geographical area:
    (1) ``How many people do you know personally in this geographical area?"
    (2) ``How many of the above have been diagnosed or have had symptoms compatible with COVID-19, to the best of your knowledge?"
The first question provides what we call the \emph{reach} $r_i$ of a participant $i$, and the second gives the number of \emph{cases} $c_i$.
The ratio $\sum_i c_i / \sum_i r_i$ provides an estimator of the fraction of the population that has had COVID-19. 

In order to estimate the
number of active cases, an additional questions has been included:
    (3) ``How many are still sick?"
So, if the answer for this question by participant $i$ is $s_i$, an estimator of the fraction of the population that is currently sick of
COVID-19 is $\sum_i s_i / \sum_i r_i$.
We label this estimate \textsf{CoronaSurveys}. 

The UMD Global CTIS, in turn, has two questions that allow respondents to report on the symptoms of their contacts: 
\begin{itemize}
    \item B3: ``Do you personally know anyone in your local community who is sick with a fever and either a cough or difficulty breathing?"
    \item B4: ``How many people do you know with these symptoms?" 
\end{itemize}
The combination of these two questions allows us to extract the number of cases of COVID-like illness $cli_i$ in the local community of a participant $i$. This information is similar to the number of sick cases $s_i$ provided by the third question above in the CoronaSurveys poll. Unfortunately,
the UMD Global CTIS does not include a question that allows us to estimate the size of the local community. Instead, we use, as an estimate of this size, the average reach $\bar{r}$ obtained with the CoronaSurveys poll in the considered geographical area\footnote{The results shown here use $\bar{r}=71$ for all countries}. Hence, we estimate the
ratio of the population that is sick with a COVID-like illness from $n$ responses to the UMD Global CTIS as 
$\sum_i cli_i /(n \cdot \bar{r})$ (labelled \textsf{UMD CLI local} in the plots).

\begin{figure}[tb]
\begin{center}
\includegraphics[width=0.9\linewidth]{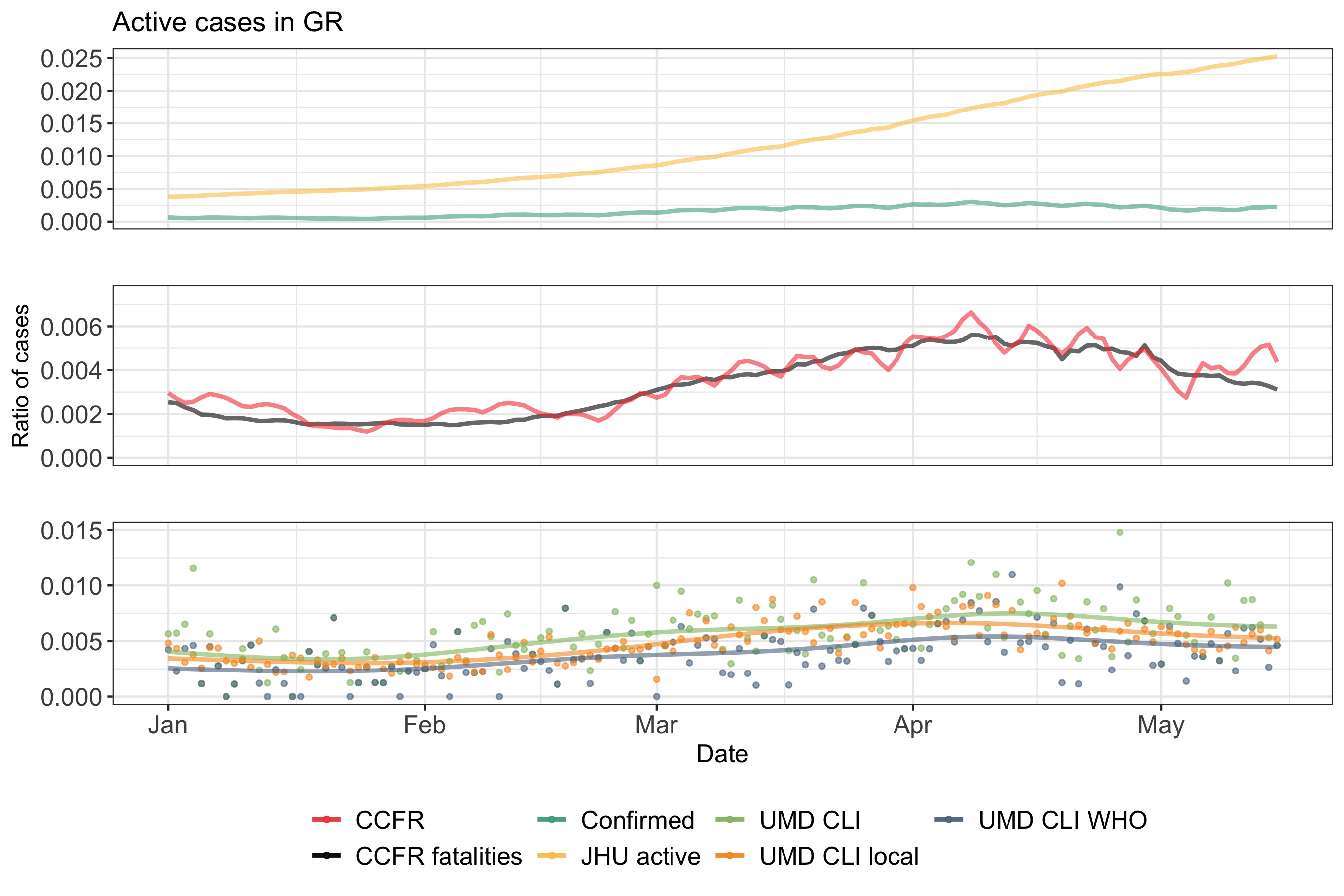}\\
\vspace*{-1cm}
\includegraphics[width=0.9\linewidth]{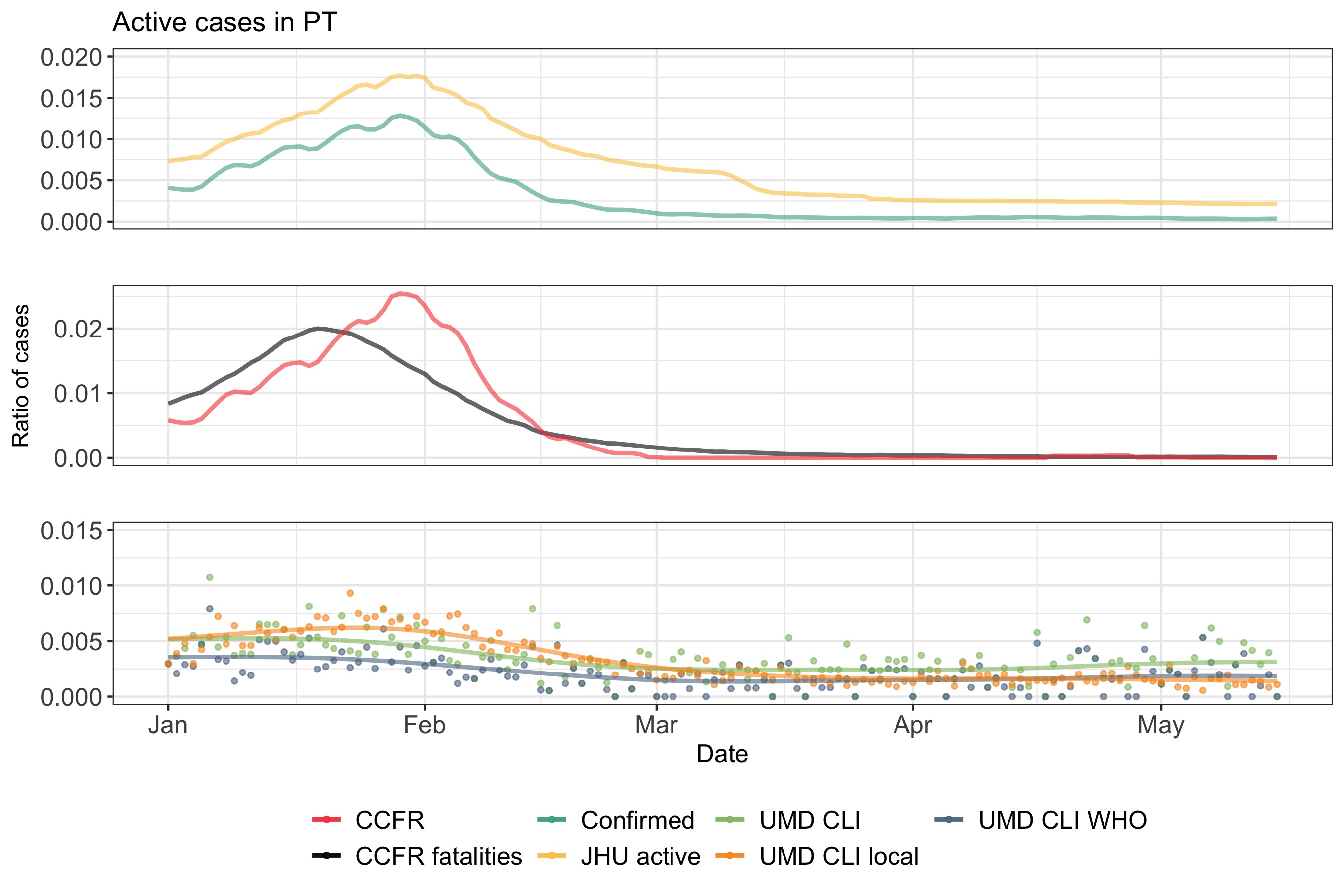}
\end{center}
\caption{Active cases in Greece and Portugal (Jan-May 2021).}
\label{f-ptgr}
\end{figure}

\section{Examples}
\label{s-examples}
\begin{figure}[tb]
\begin{center}
\includegraphics[width=0.9\linewidth]{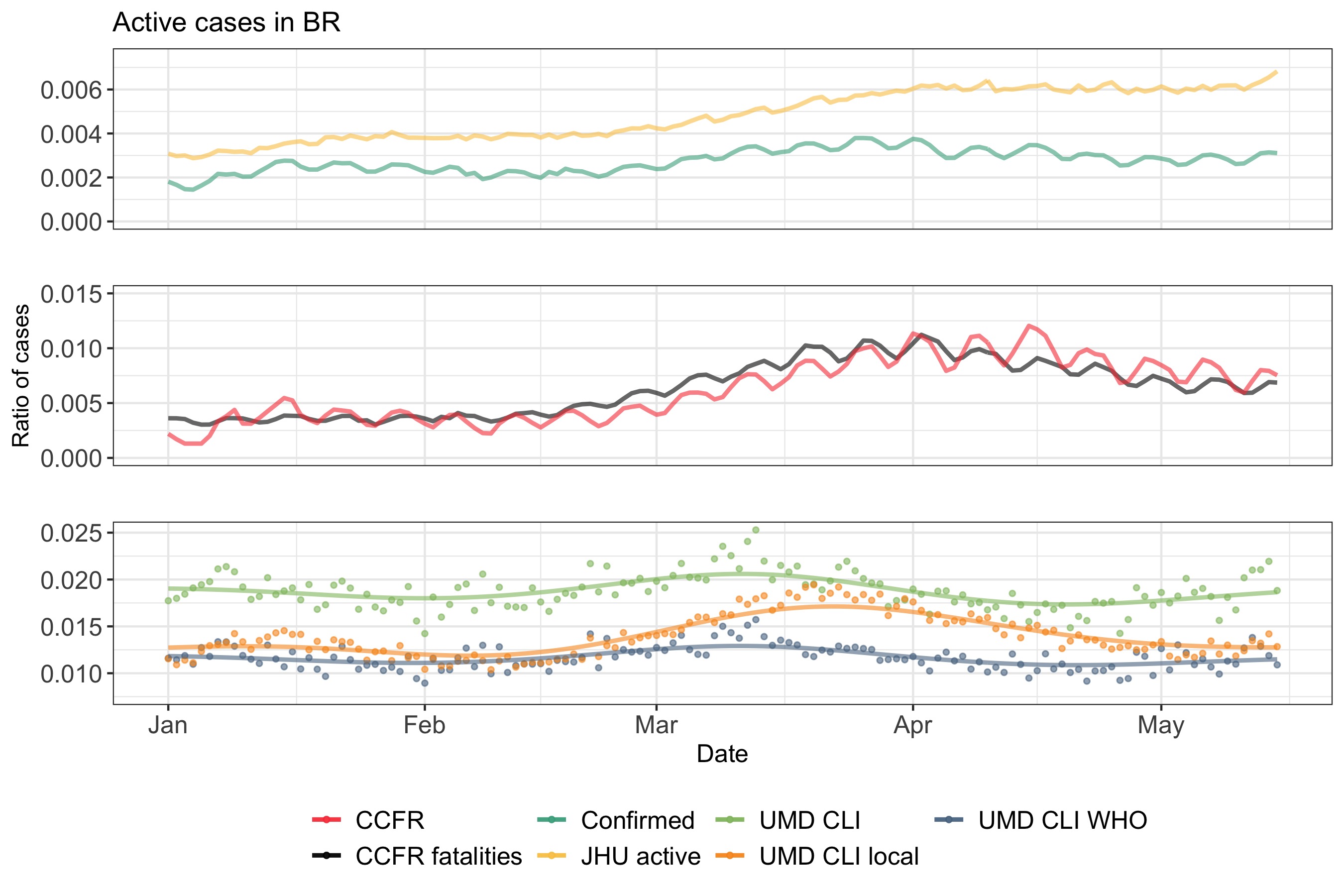}\\
\vspace*{-1cm}
\includegraphics[width=0.9\linewidth]{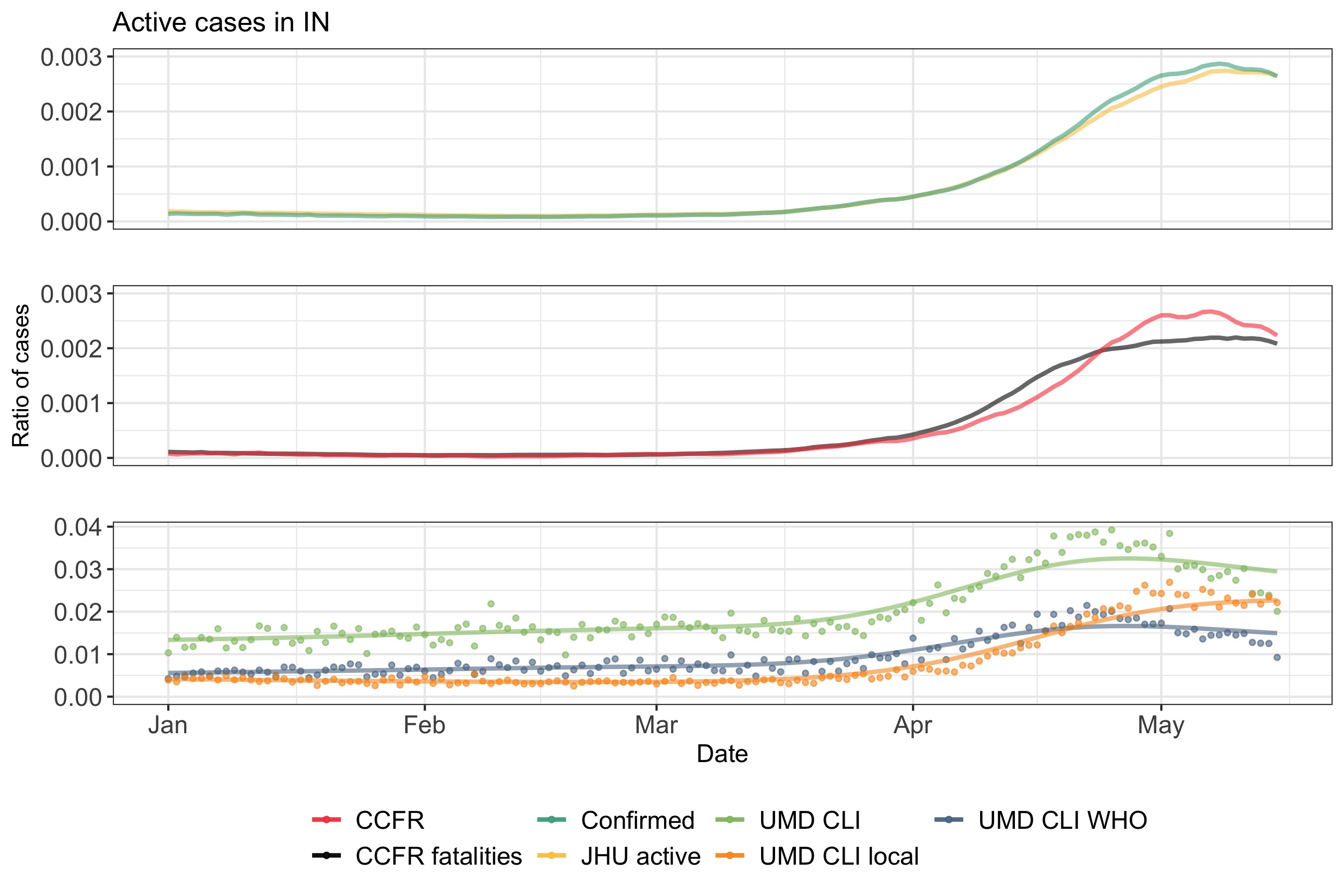}
\end{center}
\caption{Active cases in Brazil and India (Jan-May 2021).}
\label{f-ptin}
\end{figure}

In this section, we first present the estimates computed for Greece (GR), Portugal (PT), Brazil (BR), and India (IN) in 2021, reported in Figures~\ref{f-ptgr} and \ref{f-ptin}. In the general case, we observe that the estimated number of active cases using the confirmed (reported) data are lower than those derived from surveys.
For Greece, reported data may be misleading. In particular, the JHU active plot has an increasing trend on active cases while the actual confirmed cases in the country are dropping. Such behavior indicates possible issues in the number of reported cases in Greece, confirmed when comparing with \cite{worldometers}. On the other hand, the estimates based on the confirmed cases, or on the fatalities (i.e. CCFR and CCFR fatalities) appear to provide a better picture of the actual situation. Similar trends, although with slightly larger numbers, can be observed for the estimates obtained from the survey responses. 
In Portugal, we observe a high alignment between the estimates obtained by the reported cases, with the peak of the curves reaching $0.02\%$ of the population. However, the survey estimates in the case of Portugal produce a lower percentage of active cases with their peak not going above $0.005\%$ of the population. 

Brazil exhibits a different trend, where the estimates based on the fatalities are higher (almost double) than the estimates based on the confirmed cases. Such behavior may suggest that fatalities are reported better than confirmed cases. On the other hand, estimates based on surveys suggest a high under-reporting of confirmed cases, resulting in almost 10 times more active cases than what estimated by the confirmed cases. 
Finally, India seems to follow Portugal's trend when it comes to the estimates based on reported cases. The curves obtained from confirmed cases and fatalities appear to align, suggesting that the 
$\textsf{CFR}$ in the country matches the baseline $\textsf{CFR}$ from Wuhan. The estimates coming from the surveys however match the trend we observe in Brazil, suggesting that the active cases are ten times more than what estimated by the confirmed cases. Essentially this possibly indicates a significant under-reporting in the confirmed cases. 

\begin{figure}[tb]
\begin{center}
\includegraphics[width=0.9\linewidth]{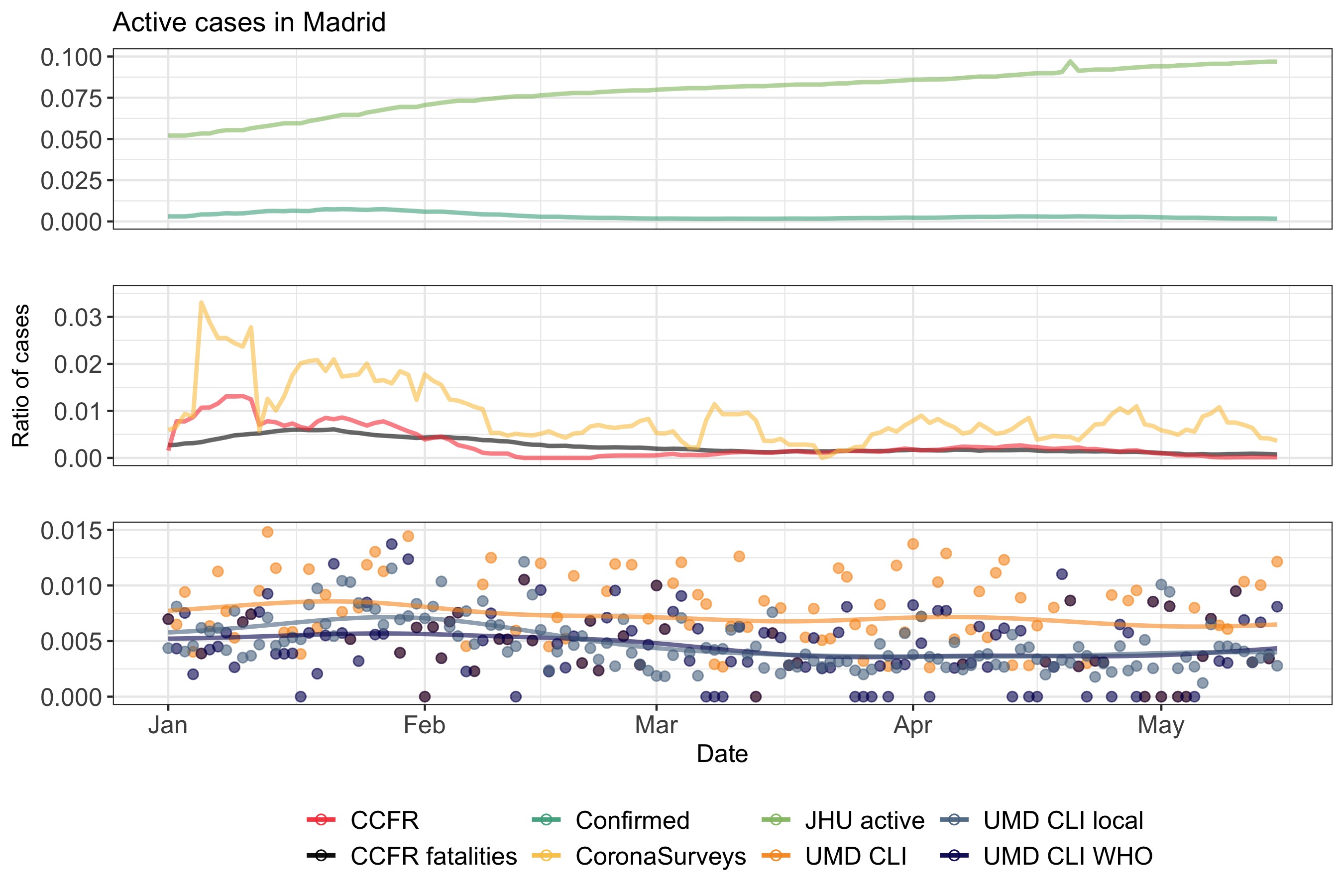}
\end{center}
\caption{Active cases in Madrid, Spain (Jan-May 2021)}
\label{f-madrid}
\end{figure}

In Figure~\ref{f-madrid} we present the estimates of active cases for Madrid, including the ones derived from the CoronaSurveys poll (which was not included in the other plots because the number of survey responses in those countries was too low). Here we see that JHU active has some misreporting problem (in fact the problem is with the number of recovered cases, like in Greece). The other estimates follow similar patterns, correctly identifying the burst of cases that Madrid suffered in January 2021. It is worth mentioning that CoronaSurveys uses an average of 36 responses per day while the other survey-based estimates have close to 500 responses per day.

\section{Discussion and Future Work}
\label{s-discussion}

The presented results have known limitations.
As mentioned, the techniques developed in Section \ref{s-fatalities} yield estimates (\textsf{CCFR} and \textsf{CCFR Fatalities}) of symptomatic cases. Moreover, \cite{Russell2020.07.07.20148460} takes as reference the baseline $\textsf{CFR}_b=1.38\%$ from Wuhan, without taking into account the differences between countries and over time \cite{sorci2020explaining,ghisolfi2020predicted}. 
Additionally, countries that have vaccination in progress typically vaccinate first the older age groups, and the consequence is that the baseline case fatality ratio will decrease, since younger groups have lower CFR \cite{sorci2020explaining}.
We are exploring how to adapt these techniques to obtain estimates of infected cases, using for that the Infection Fatality Rate ($\textsf{IFR}$), and taking into account country, age \cite{o2021age}, vaccination, and SARS-CoV-2 variants. 
Another limitation is the use of a fixed period (median value) for a case to be active and for onset to death, which we plan to replace with random variables. 
Regarding survey-based estimates, we are exploring how to improve the curating of the responses and how to take into account the selection bias of the surveys. Specifically for indirect reporting, we are exploring the influence of the social network structure and how to have better average reach estimates. Additionally, we are exploring have to deal with the known biases of NSUM \cite{laga2021thirty}.

\begin{figure}[tb]
\begin{center}
\includegraphics[width=0.3\linewidth]{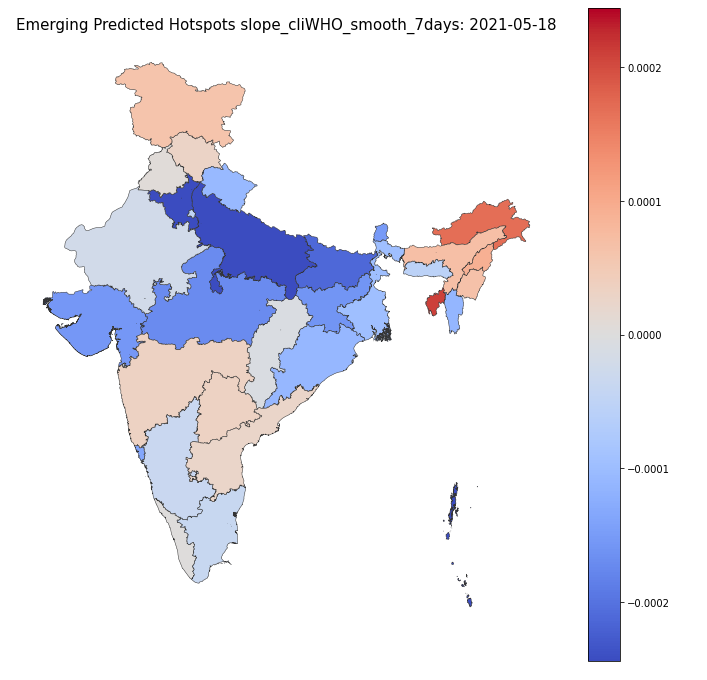}~~~~~~~~
\includegraphics[width=0.3\linewidth]{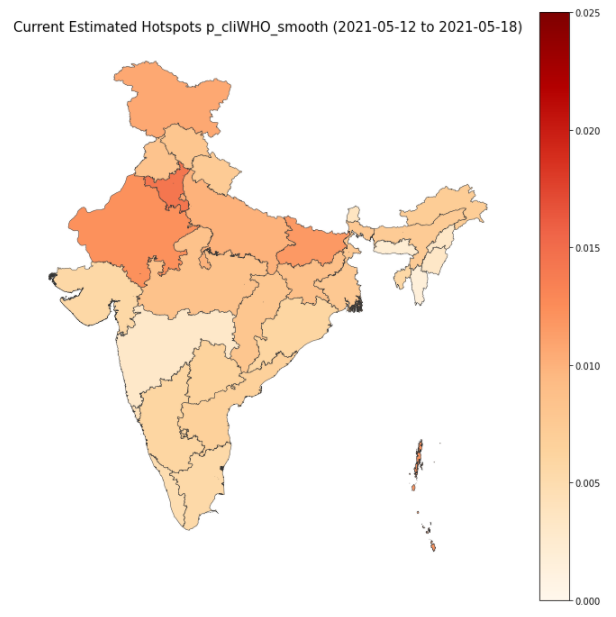}~~~~~~~~
\includegraphics[width=0.4\linewidth]{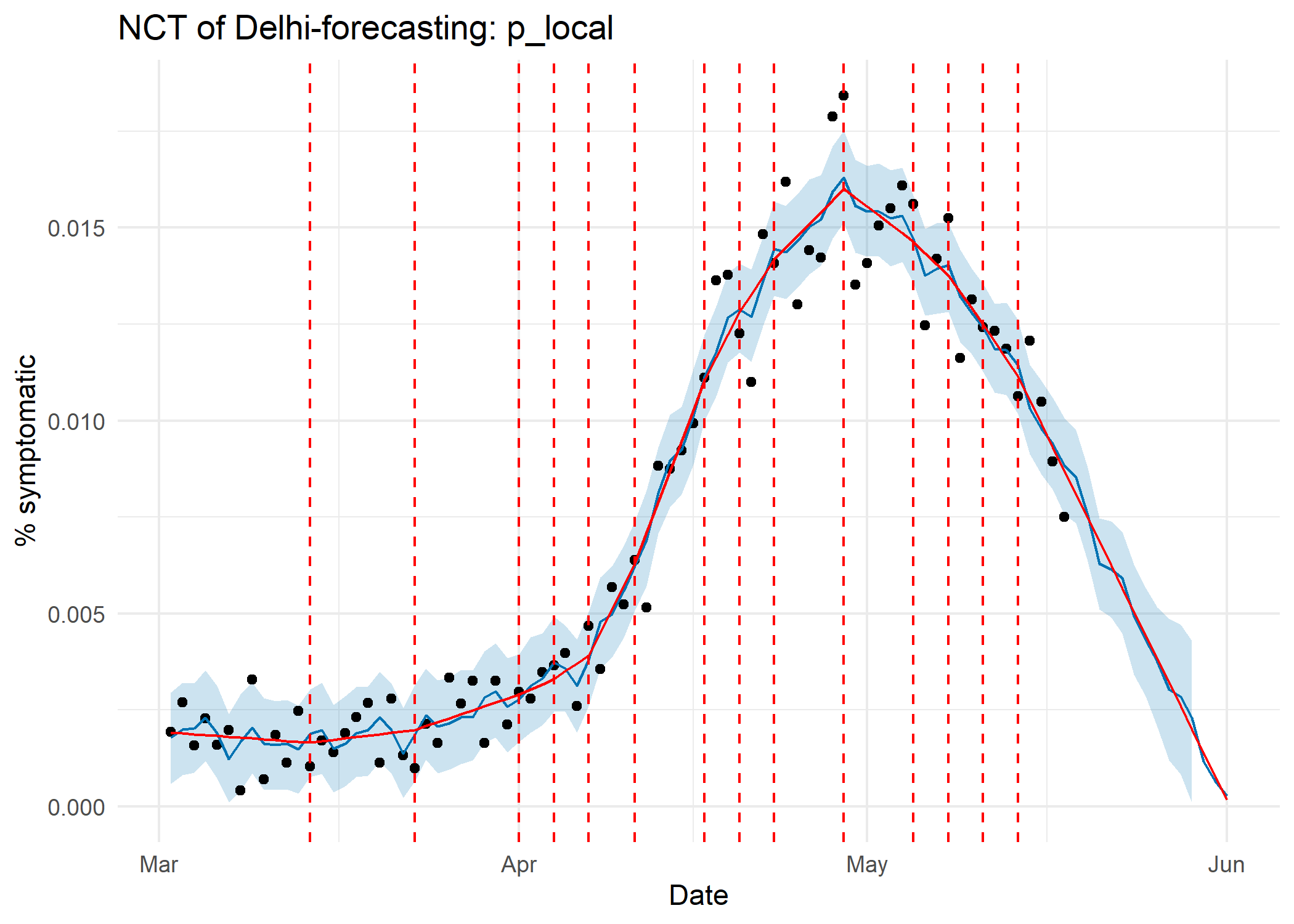}
\end{center}
\caption{Maps of current ratio of active cases (left) and its past evolution (center) in India. Forecasting of active cases in Delhi, India (right).}
\label{f-prophet}
\end{figure}

As future work, we will be using these estimates to help authorities understand the real situation of the pandemic in each region and its most likely evolution in the future. This will enable the identification of future hot areas and prepare them for the rise in cases. For this we are producing maps and charts that show the current situation in terms of active cases at regional level, and that show how this situation has changed in the recent few days (see Figure~\ref{f-prophet}, left and center)
Along similar lines, we will use time-series analysis to forecast the evolution of the number of active cases in the near future (see Figure~\ref{f-prophet}, right). 

Finally, we are developing a data-driven approach to process the UMD Global CTIS data to determine the probability that a response is an active case of COVID-19 from the set of symptoms it declares. To this end, we will use the responses that give
a positive answer to the question ``Have you been tested for coronavirus (COVID-19) in the last 14 days?" as training data, assuming that the result of that test tells us whether the respondent is an active COVID-19 case.

\section*{Acknowledgements}

We want to thank the whole CoronaSurveys Team \cite{coronasurveys-web} for the collective effort. We also want to thank Tamer Farag and Kris Barkume for useful discussions on the survey data. This work is partially supported by grant CoronaSurveys-CM, funded by IMDEA networks and Comunidad de Madrid.

\bibliographystyle{plain}
\bibliography{refs}

\end{document}